\documentclass[conference]{IEEEtran}
\IEEEoverridecommandlockouts
\usepackage{cite}
\usepackage{amsmath,amssymb,amsfonts}
\usepackage{algorithmic}
\usepackage{float}
\usepackage{graphicx}
\usepackage{textcomp}
\usepackage[dvipsnames]{xcolor,colortbl}
\usepackage{booktabs}
\usepackage{arydshln}
\usepackage{microtype}
\usepackage{cleveref}
\def\BibTeX{{\rm B\kern-.05em{\sc i\kern-.025em b}\kern-.08em
    T\kern-.1667em\lower.7ex\hbox{E}\kern-.125emX}}

\renewcommand{\newline}{\vspace{0.5em}}
\newcommand{\ie}{\textit{i.e.}}
\newcommand{\eg}{\textit{e.g.}}

\begin{document}

\title{SAGE: Semantic-Driven Adaptive Gaussian\\Splatting in Extended Reality}

\author{\IEEEauthorblockN{Chiara Schiavo, Elena Camuffo, Leonardo Badia and Simone Milani}
\IEEEauthorblockA{Dept. of Information Engineering, University of Padova, Padua, Italy\\}
Emails: \small{\texttt{\{chiara.schiavo, elena.camuffo, leonardo.badia, simone.milani\}@unipd.it}}
}

\maketitle

\begin{abstract}
3D Gaussian Splatting (3DGS) has significantly improved the efficiency and realism of three-dimensional scene visualization in several applications, ranging from robotics to eXtended Reality (XR). This work presents SAGE (\underline{S}emantic-Driven \underline{A}daptive \underline{G}aussian Splatting in \underline{E}xtended Reality), a novel framework designed to enhance the user experience by dynamically adapting the Level of Detail (LOD) of different 3DGS objects identified via a semantic segmentation. Experimental results demonstrate how SAGE effectively reduces memory and computational overhead while keeping a desired target visual quality, thus providing a powerful optimization for interactive XR applications.
\end{abstract}


\section{Introduction}
\label{sec:intro}

The rapid evolution of eXtended Reality technologies on mobile and wearable platforms has increased the demand for efficient 3D rendering techniques that provide highly immersive and fluid user experiences. However, balancing computational efficiency and visual quality introduces several challenges, particularly in resource-constrained environments~\cite{cap19:pcc,jav21:qual}. 
Traditional approaches often rely on geometric simplification or predefined Level of Detail strategies, which are chosen according to user proximity and interaction~\cite{Bil14:lod,Has21:3Dsimple}. Previous research has focused on adapting cognitive load~\cite{dos14:ar_cognitive}, predicting user actions to optimize training experiences~\cite{VAUGHAN201665}, or minimizing transmitted information~\cite{4351598}, while only a few recent attempts have been made using Deep Neural Networks~\cite{camuffo2022deep}. The advent of neural representation techniques, such as Neural Radiance Fields (NeRF)~\cite{nerf} and 3D Gaussian Splatting~\cite{kerbl3Dgaussians}, has transformed 3D scene rendering. These methods enable implicit scene representations that maintain high fidelity while offering flexibility for on-demand rendering. While NeRF emphasizes detailed and photorealistic scenes using a single neural network, 3DGS leverages Gaussian-shaped primitives for lightweight yet differentiable scene management and rendering.

In this paper, we introduce SAGE (\underline{S}emantically \underline{A}daptive \underline{G}aussian Splatting in \underline{E}xtended Reality), a novel approach that integrates semantic information in the optimization process of 3DGS. By performing semantic segmentation, SAGE dynamically adjusts the 
3DGS representation quality of individual scene components based on their spatial and visual importance. This method leads to a reduction in memory and computational overhead while maintaining high visual quality. Through extensive evaluations on different scenes of the Mip-NeRF360 dataset~\cite{mipnerf360}, we demonstrate SAGE’s ability to significantly enhance efficiency and scalability in 3DGS rendering. 
This ability proves to be extremely suitable in XR applications, where rendering complex scenes in real-time is essential, and selecting quality based on scene semantics can be beneficial to ensure a better user experience.
SAGE is designed to optimize resource allocation rather than maximize perceptual quality alone, maintaining comparable visual quality with significantly lower memory usage compared to other standard approaches.

\section{Related Work}
\label{sec:related}

\begin{figure*}[t]
    \centering
    \includegraphics[width=\linewidth]{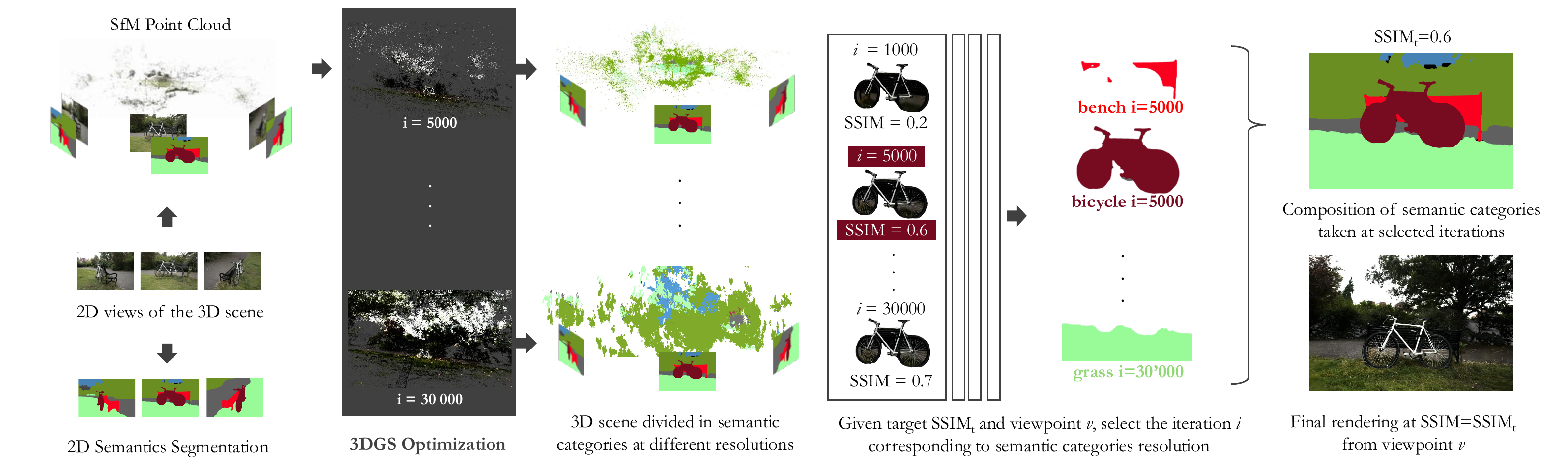}
    \caption{SAGE pipeline. Starting from a set of 2D views \( V \), SAGE retrieves the 2D semantics using DeepLabV2~\cite{deeplabv2}. In parallel, it constructs the Structure from Motion point cloud, like in standard 3DGS. Then it processes the SfM point cloud at increasing resolution with proceeding iteration $i$. Differently from standard 3DGS, SAGE follows the semantic masks provided on 2D views to partition the 3D point cloud and perform selective optimization of different semantic categories. By setting a target quality (SSIM$_t$) the optimization of each semantic category stops the optimization process when such target value is achieved. The final render from selected viewpoint $v$ is obtained as a composition of the scene categories optimized separately for target quality. }
    \label{fig:pipeline}
\end{figure*}

The problem of quality maximization under bandwidth constraints first appeared in image, audio and video compression~\cite{6178830}. Later, several techniques in networking %
were proposed for HTTP Adaptive Streaming (HAS) to modulate the data stream while minimizing the quality decrement~\cite{Wang17:qoe_has,10.1145/3488660.3493802}.
Recently, deep learning solutions were investigated to address this problem in the video domain~\cite{9122485} %
and allowed the extension to XR applications.
Whenever referring to XR applications, visual quality does not rely only on the modeling accuracy for the synthetic 3D objects, but also on the rendering complexity and smoothness as viewpoint changes.

With the advent of 3DGS~\cite{kerbl3Dgaussians} as a powerful technique for real-time rendering, it become possible to represent 3D scenes via Gaussian primitives in a more flexible manner, even if the representation of high-quality scenes remains computationally demanding due to the large number of Gaussians required. To address these issues, several techniques have been proposed so far.
LightGaussians~\cite{lightgaussian} and Compact3D~\cite{C3DGS} use pruning and quantization to reduce memory, while  Multi-Scale 3DGS~\cite{multiscale3dgaussiansplatting} introduces multi-scale representations to enhance fidelity and mitigate aliasing. OctreeGS~\cite{octreegs} employ hierarchical structures to optimize memory and computation, while FlOD~\cite{flod} applies a Level of Detail strategy for real-time rendering based on device constraints. 

Despite these improvements, experimental results have shown that tayloring the computational effort to the specific objects allows a better complexity saving while preserving the visual quality. To this purpose, SAGE identifies the different objects by means of sematic segmentation, which has also been investigated
in 3D models reconstruction \cite{campagnolo2023fully} 
or transmission~\cite{mari2023cactus}.
Recently, a few attempts have also been made to integrate semantic understanding tasks within 3DGS~\cite{schieber2024semanticscontrolledgaussiansplattingoutdoor} to jointly improve understanding and reconstruction.
\section{Methodology}
\label{sec:method}

\paragraph{Preliminaries} %
3D Gaussian Splatting~\cite{kerbl3Dgaussians} is an explicit radiance field technique for efficient, high-quality rendering. It represents a scene using differentiable 3D Gaussian primitives optimized to fit the scene's geometry, capturing shapes and features in a compact, expressive form. Given a set of images, 3DGS is initialized from a point cloud generated by a Structure from Motion (SfM) algorithm, and then iteratively refined through adaptive point densification and pruning, enabling high-quality rendering.

\newline
\paragraph{Proposed Approach} SAGE utilizes semantic segmentation to increase 3DGS optimization performance on individual objects and develop an adaptive quality system.
Semantic segmentation is applied to the input 2D views~\cite{deeplabv2} assigning a label \( l \in L \) to each pixel in every image \( v \in V \). These labels are then related to 3D points in the SfM reconstruction via the estimated camera parameters, using a majority voting scheme that resolves conflicts by assigning each 3D point the most frequent label across its projections.

To perform the optimization, we iteratively compute pre-defined quality metrics, such as PSNR and SSIM, for each fixed viewpoint (rendered image) $v$ and level of reconstruction $i$, restricted to every semantic label $l$ separately. These quality metrics are computed with respect to $d_{min,l,v}$, which represents the distance of the closest point assigned to semantic label \(l\) with respect to the camera position, to be conservative in the optimization process.
Formally, given a fixed viewpoint $v = \omega$, the quality-constrained optimization problem is designed as follows:
\begin{equation}\small
\min_{i} \sum_l N_l(i) \quad \text{s.t.} \quad \text{SSIM}_{l,i}(d_{min,l}) \geq \text{SSIM}_t, \ \forall l\in L
\end{equation}
where $i$ represents the minimum iteration of the 3DGS algorithm to ensure the desired quality (with $\text{SSIM}_t$ denoting the target SSIM value) is met for the semantic category $l$, while minimizing the number of Gaussian primitives used. 
$N_l(i)$ denotes the number of Gaussians representing category $l$ at iteration $i$. The quality at iteration $i$ is represented by $\text{SSIM}_{l,i}$ and depends on the distance between the position of the camera and the closest 3D point to the camera, with label $l$, {\ie}, $d_{min,l} = d_{min,l,v}|_{v=\omega}$ with a fixed viewpoint.
As a result, a parametric model can be fit for a target $\text{SSIM}_{t}$:
\begin{equation}\small
   \text{SSIM}_{l,i}(d_{min,l}) =
\begin{cases} 
    K_1 \cdot e^{- \gamma_1 |d_{min,l} -\mu_1|^{\alpha_1}}  & \text{if } d_{min,l} < \beta, \\
    K_2 \cdot e^{- \gamma_2 |d_{min,l} -\mu_2|^{\alpha_2}}  & \text{if } d_{min,l} \geq \beta;
\end{cases}
\label{eq:fitting}
\end{equation}
where the coefficients $K_n$, $\gamma_n$, $\mu_n$, and $\alpha_n$ (with $n = 1, 2$) and the threshold $\beta$ that separates the two distance regimes are obtained by fitting the equation to the desired label $l$ at the level of reconstruction detail $i$ (see Sec.~\ref{sec:results} and Fig.~\ref{fig:fitting} for clarity). This way, the model is able to predict for a target class $l$ at which iteration $i$ the 3DGS algorithm should be halted to meet the desired quality $\text{SSIM}_{t}$. The operation is iterated on all viewpoints $v \in V$, obtaining distance-dependent fitting. An overall representation of SAGE is shown in Fig.~\ref{fig:pipeline}.

Note that this technique is efficient in terms of computational efficiency as it reallocates computational resources based on semantic importance, while keeping visual quality sufficiently high. This makes the semantic-based optimization of SAGE applicable to many rendering systems, in addition to 3DGS.

\newline
\paragraph{Evaluation methodologies} In order to measure both the adaptability and robustness of the proposed approach, two types of evaluation tests were performed: (i) cross-view and (ii) cross-scene. 
The (i) {cross-view} test focuses on synthesizing a novel view starting from existing views of a single 3D scene using the model for fitting different labels. %
The 3D scene is partitioned into class-labeled 3D points by leveraging the segmentation of 2D views and projecting onto the 3D point cloud. Semantic masks are extracted by projecting back the semantics from the 3D points to 2D onto the novel view, and the model is optimized on the novel view according to the desired visual quality.
Once the new view has been synthesized, the SSIM is evaluated on two levels: first, for the entire view to check overall image quality, and second, for each semantic mask to verify that the prediction is correct. 
The (ii) {cross-scene} test examines whether a model trained on a semantic category (\eg, \textit{grass-merged} from ``bicycle'' scene) generalizes to the same category in another scene (\eg, \textit{grass-merged} from ``garden'' scene). If the SSIM curve remains consistent, the learned parameters and iteration values are transferable, enabling a general optimization routine applicable to unseen data.

\section{Experimental results}
\label{sec:results}
\begin{table}[t]
    \centering \scriptsize
    \setlength{\tabcolsep}{0.2em}
    \renewcommand{\arraystretch}{1}
        \caption{Average results of SAGE for scene ``bicycle".}%
    \label{avg-bicycle}
    \begin{tabular}{l|rr:rr:rr:rr}
    \toprule
   \multicolumn{1}{c}{} &  \multicolumn{2}{c}{5000} & \multicolumn{2}{c}{10'000} & \multicolumn{2}{c}{15'000} & \multicolumn{2}{c}{30'000}\\
   \cmidrule(lr){2-3} \cmidrule(lr){4-5} \cmidrule(lr){6-7} \cmidrule(lr){8-9} \rm
     & SSIM & \# Gauss. & SSIM & \# Gauss. & SSIM & \# Gauss. &  SSIM & \# Gauss.  \\
     \midrule
    Bench & $0.598$ & $141$k & $0.635$ & $234$k & $0.659$ & $334$k & $0.682$ & $336$k\\
    Bicycle & $0.602$ & $50$k & $0.658$ & $111$k & $0.674$ & $144$k & $0.688$ & $146$k \\
    \rowcolor{CadetBlue!15} Grass-merged & $0.438$ & $287$k & $0.491$ & $703$k & $0.521$ & $972$k & $0.537$ & $985$k \\
    \rowcolor{CadetBlue!15} Pavement-merged & $0.585$ & $163$k & $0.625$ & $337$k & $0.642 $& $422$k & $0.651$ & $424$k\\
    Sky-other-merged & $0.917$ & $278$k & $0.920$ & $456$k & $0.923 $& $587$k & $0.923$ & $578$k \\
    \rowcolor{CadetBlue!15}Tree-merged & $0.540$  & $1.4$M & $0.574$ & $2.5$M & $0.594$ & $3.2$M & $0.613$ & $3.3$M \\
    \midrule
    Total & $0.546$ & $2.3$M & $0.598$ & $4.3$M & $0.627$ & $5.7$M & $0.647$ & $5.8$M\\
    \bottomrule
    \end{tabular}
\end{table}

\begin{table}[t]
    \centering \scriptsize
    \setlength{\tabcolsep}{0.2em}
    \renewcommand{\arraystretch}{1}
        \caption{Average results of SAGE for scene ``garden".}
    \label{avg-garden}
    \begin{tabular}{l|rr:rr:rr:rr}
    \toprule
   \multicolumn{1}{c}{} &  \multicolumn{2}{c}{5000} & \multicolumn{2}{c}{10'000} & \multicolumn{2}{c}{15'000} & \multicolumn{2}{c}{30'000}\\
   \cmidrule(lr){2-3} \cmidrule(lr){4-5} \cmidrule(lr){6-7} \cmidrule(lr){8-9} \rm
     & SSIM & \# Gauss. & SSIM & \# Gauss. & SSIM & \# Gauss. &  SSIM & \# Gauss.  \\
     \midrule
    Dining table & $0.828$ & $211$k & $0.869$ & $309$k & $0.879$ & $379$k & $0.880$ & $379$k \\
    \rowcolor{CadetBlue!15} Grass-merged & $0.691$ & $742$k & $0.776$ & $1.1$M & $0.798$ & $1.3$M & $0.804$ & $1.3$M \\
    \rowcolor{CadetBlue!15} Pavement-merged & $0.753$ & $447$k & $0.805$ & $682$k & $0.816$ & $736$k & $0.818$ & $736$k \\
    Potted plant & $0.765$ & $56$k & $0.808$ & $82$k & $0.827$ & $94$k & $0.829$ & $94$k \\
    \rowcolor{CadetBlue!15}Tree-merged & $0.688$ & $1.2$M & $0.739$ & $2.0$M & $0.749$ & $2.3$M & $0.757$ & $2.3$M \\
    Vase & $0.847$ & $14$k & $0.892$ & $17$k & $0.902$ & $18$k & $0.906$ & $18$k \\
    \midrule
    Total & $0.778$ & $3.1$M & $0.826$ & $4.9$M & $0.841$ & $5.6$M & $0.848$ & $5.6$M \\
    \bottomrule
    \end{tabular}
\end{table}

\begin{figure}[t]
    \centering
    \includegraphics[width=\linewidth]{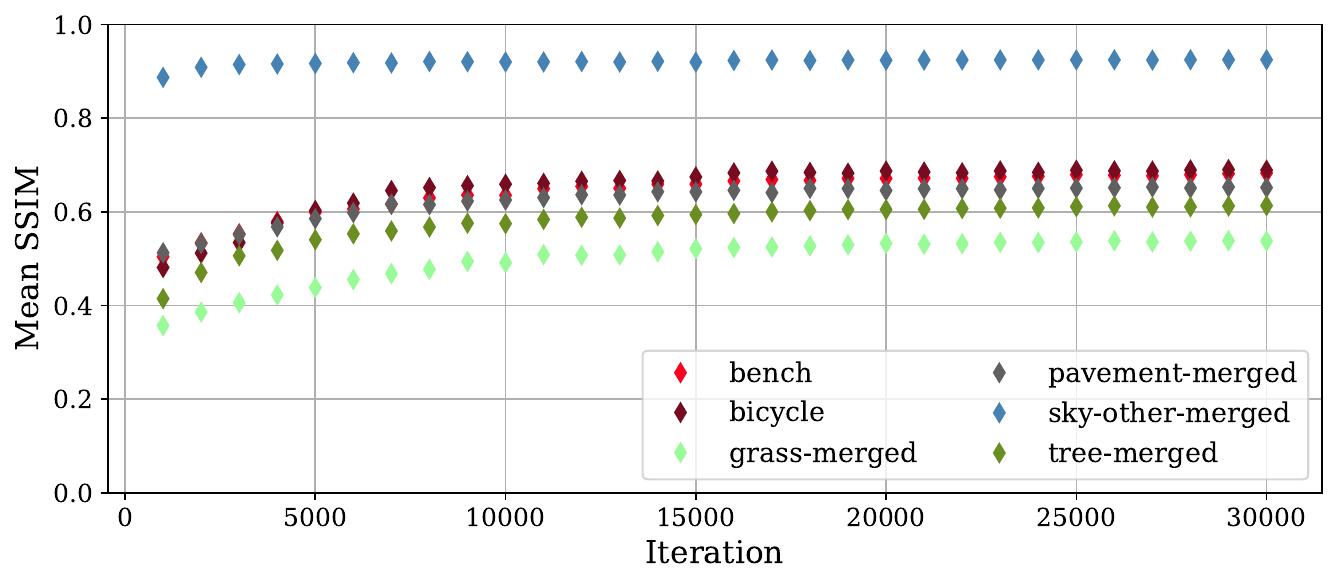} %
    \caption{Mean SSIM over training iterations for individual scene components of scene ``bicycle". Variations in optimization performance across semantic categories are visible. Highly textured content (\eg, \textit{grass-merged}) show lower overall quality compared to smooth areas (\eg, \textit{sky-other-merged}).}
    \label{fig:ssim-iter} %
\end{figure}

\begin{figure*}[t]
    \centering
    \begin{minipage}{0.33\textwidth}
    \centering
    \includegraphics[width=1\textwidth]{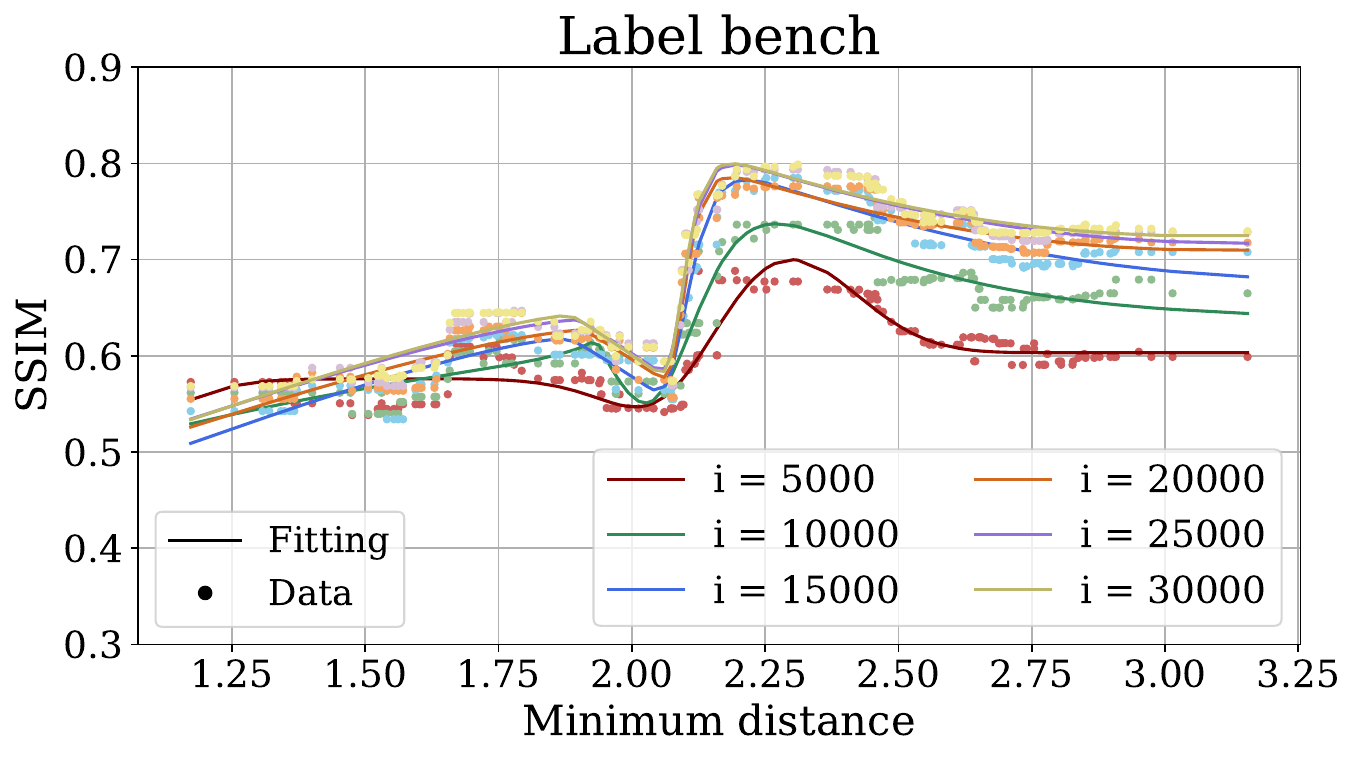}
    \end{minipage}%
    \begin{minipage}{0.33\textwidth}
    \centering
    \includegraphics[width=1\textwidth]{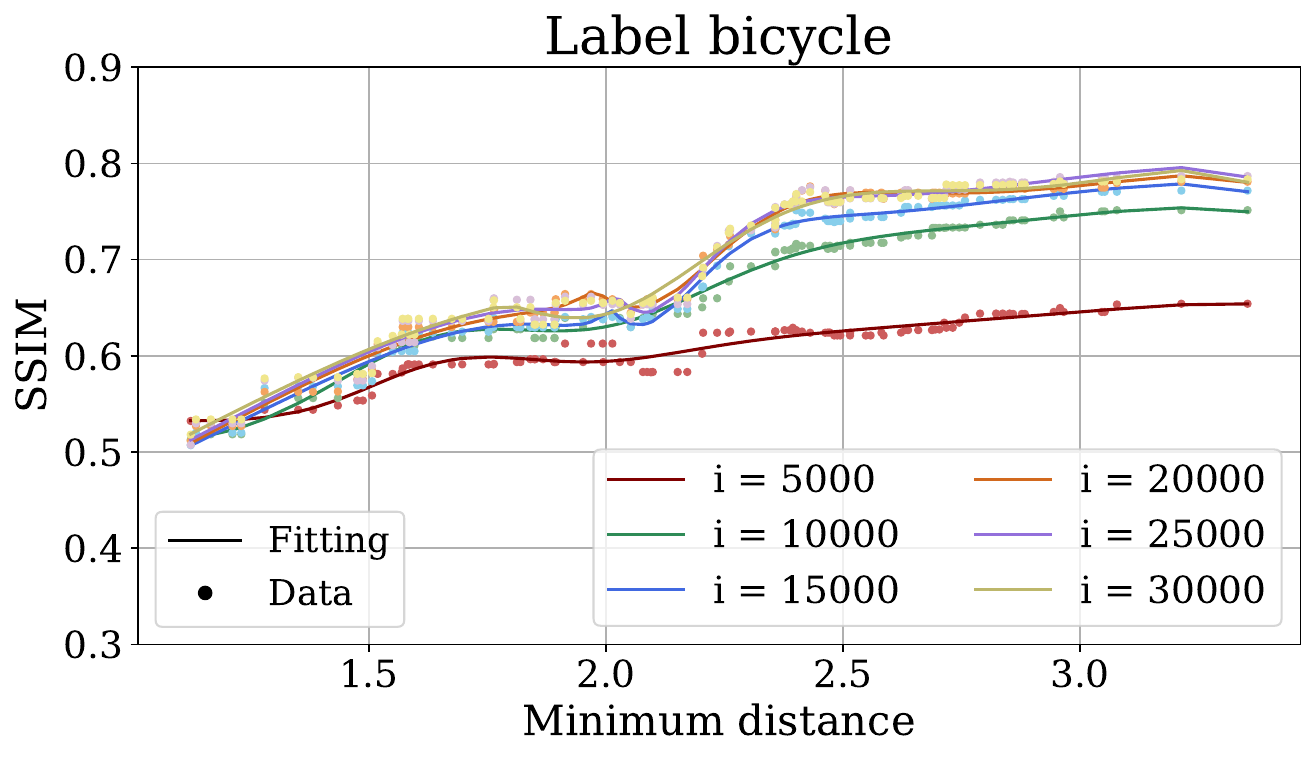}
    \end{minipage} %
    \begin{minipage}{0.33\textwidth}
    \centering
    \includegraphics[width=1\textwidth]{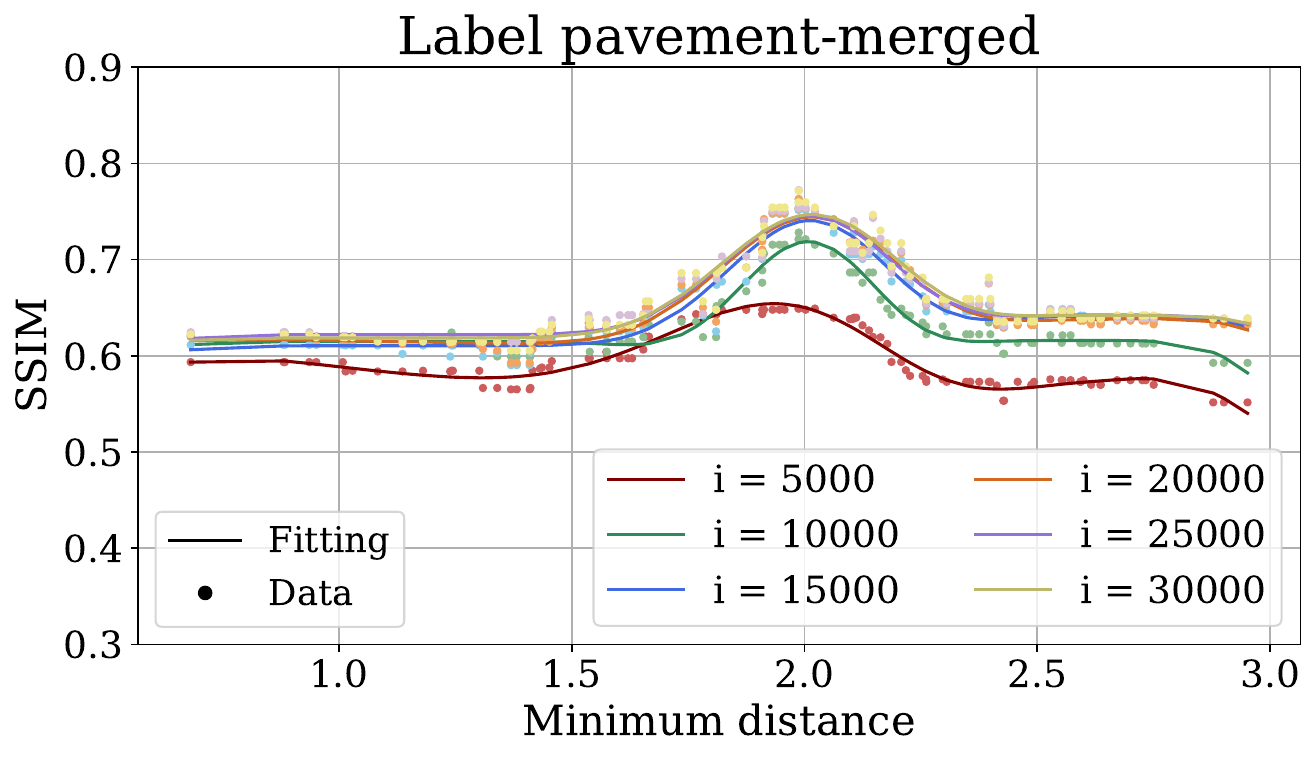}
    \end{minipage} 
    
    \caption{SSIM as a function of the minimum distance for the semantic labels \textit{bench}, \textit{bicycle} and \textit{pavement-merged}. Each curve represents data collected at a different iteration $i$ of 3DGS, with experimental data (dots) and fitted trends (lines). The SSIM generally increases with distance, reaching a peak before stabilizing or declining, with higher iterations showing improved reconstruction quality and smoother trends.}
    \label{fig:fitting}
\end{figure*}

\begin{table*}[t]
    \centering \scriptsize
        \setlength{\tabcolsep}{0.15em}
    \renewcommand{\arraystretch}{1}
        \caption{Quantitative results on single view ``{\rm \texttt{DSC8719}}" of scene ``bicycle".}
    \label{tab:single-view}
    \begin{tabular}{l|rr:rrrr:rrrr:rrrr}
    \toprule
    \multicolumn{1}{c}{} & \multicolumn{2}{c}{Distance} & \multicolumn{4}{c}{SSIM$_i$} & \multicolumn{4}{c}{\# Gaussians} & \multicolumn{4}{c}{Occupancy $\downarrow$} \\
    \cmidrule(lr){2-3} \cmidrule(lr){4-7} \cmidrule(lr){8-11} \cmidrule(lr){12-15} 
     & Min & Avg & $5000$ & $10'000$ & $15'000$ & $30'000$ (\tiny{3DGS}) & 3DGS & SAGE$_{t = 0.5}$ & SAGE$_{t = 0.6}$ & SAGE$_{t = 0.7}$ & 3DGS & SAGE$_{t = 0.5}$ &  SAGE$_{t = 0.6}$ & SAGE$_{t = 0.7}$ \\
    \midrule \rm
    Bench & $2.615$ & $4.621$ & \cellcolor{Goldenrod!70} $0.587$ & $0.697$ &  \cellcolor{Apricot!60} $0.742$ & $0.750$ &  $336'632$ & $141'804$ & $141'804$ & $334'433$ & $83.5$ MB & $35.2$ MB & $35.2$ MB & $82.9$ MB\\
    Bicycle & $3.046$ & $4.358$ & \cellcolor{Goldenrod!70} $0.632$ &  \cellcolor{Apricot!60} $0.713$ & $0.759$ & $0.758$ &  $146'103$ & $50'965$ & $50'965$ & $111'799$ & $36.2$ MB & $12.6$ MB & $12.6$ MB & $27.7$ MB\\
    Grass-merged & $1.609$ & $6.809$ & $0.341$ & $0.393$ & $0.434$ &  \cellcolor{Apricot!60}$0.419$ & $985'863$ & $985'863$ & $985'863$ & $985'863$ & $224.5$ MB & $224.5$ MB & $224.5$ MB & $224.5$ MB\\
    Pavement-merged & $2.671$ & $6.335$ & \cellcolor{YellowGreen!50} $0.586$ & \cellcolor{Goldenrod!70} $0.662$ & $0.688$ &  \cellcolor{Apricot!60} $0.693$ &  $424'146$ & $163'298$ & $337'966$ & $424'124$ & $105.2$ MB & $40.5$ MB & $83.8$ MB & $105.2$ MB \\
    Sky-other-merged & $8.980$ & $27.192$ &  \cellcolor{Apricot!60} $0.805$ & $0.794$ & $0.795$ & $0.803$ &  $578'378$ & $278'139$ & $278'139$ & $278'139$ & $143.4$ MB & $69.0$ MB & $69.0$ MB & $69.0$ MB\\
    Tree-merged & $0.753$ & $16.424$ & \cellcolor{YellowGreen!50} $0.595$ & $0.612$ &  \cellcolor{Apricot!60} $0.642$ & $0.634$ &  $3'343'641$ & $1'428'984$ & $3'343'635$ & $3'343'641$ & $829.2$ MB & $354.4$ MB & $829.2$ MB & $829.2$ MB \\
    \midrule
    Total & \multicolumn{1}{c}{|} & \multicolumn{1}{c:}{|} & $0.531$ & $0.579$ & $0.615$ & $0.617$ & $5'832'994$ & $3'049'053$ & $5'138'872$ & $5'477'999$ &$1.45$ GB & \cellcolor{YellowGreen!50}$756.2$ MB & \cellcolor{Goldenrod!70}$1,27$ GB & \cellcolor{Apricot!60}$ 1.36$ GB\\
    \bottomrule
    \end{tabular}
    \label{tab:view-dep}
\end{table*}

\begin{table}[t]
    \centering \scriptsize
    \setlength{\tabcolsep}{0.7em}
    \renewcommand{\arraystretch}{1}
        \caption{3DGS and SAGE comparison at fixed occupancy in ``bicycle".}
    \label{tab:occupancy}
    \begin{tabular}{l|rrr:rrr}
    \toprule
   \multicolumn{1}{c}{} &  \multicolumn{3}{c}{3DGS} & \multicolumn{3}{c}{SAGE} \\
   \cmidrule(lr){2-4} \cmidrule(lr){5-7} \rm
    Occupancy & SSIM $\uparrow$ & PSNR $\uparrow$ & LPIPS $\downarrow$  & SSIM $\uparrow$ & PSNR $\uparrow$ & LPIPS $\downarrow$ \\
     \midrule
     $\sim$ $700$ MB & $0.533$ & $22.057$ & $0.38$ & \cellcolor{CadetBlue!15}$0.557$ & \cellcolor{CadetBlue!15}$22.497$ & \cellcolor{CadetBlue!15} $0.27$\\
     $\sim$ $1,25$ GB & \cellcolor{CadetBlue!15}$0.599$ & $22.594 $ & \cellcolor{CadetBlue!15} $0.24$ & $0.581$ & \cellcolor{CadetBlue!15} $22.669$ & \cellcolor{CadetBlue!15} $0.24$\\
     $\sim$ $1,3$ GB & $0.612$ & $22.629$ & $0.23$ & \cellcolor{CadetBlue!15} $0.618$ & \cellcolor{CadetBlue!15} $22.816$ & \cellcolor{CadetBlue!15} $0.19$\\
    \bottomrule
    \end{tabular} %
\end{table}

\begin{figure*}
    \centering
    \includegraphics[width=\linewidth]{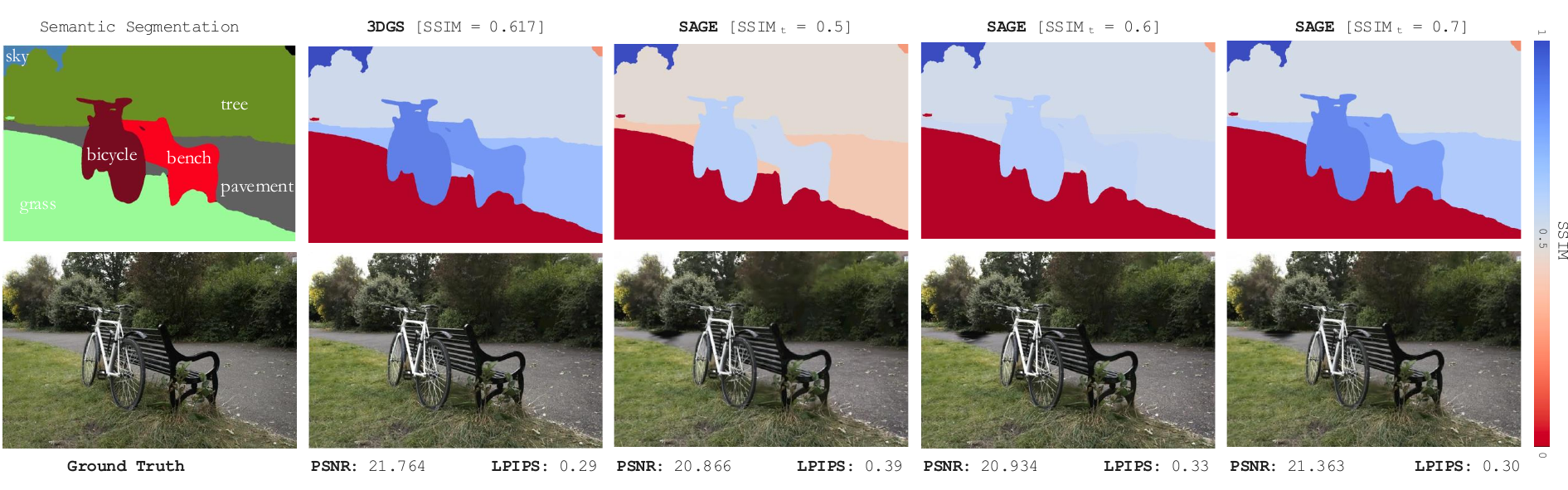} \vskip -1ex
    \caption{Qualitative results on view ``{\rm \texttt{DSC8719}}" of scene ``bicycle".} 
    \label{fig:qualitative-bicycle} \vskip -1.5ex
\end{figure*}
\begin{figure*}
    \centering
    \includegraphics[width=\linewidth]{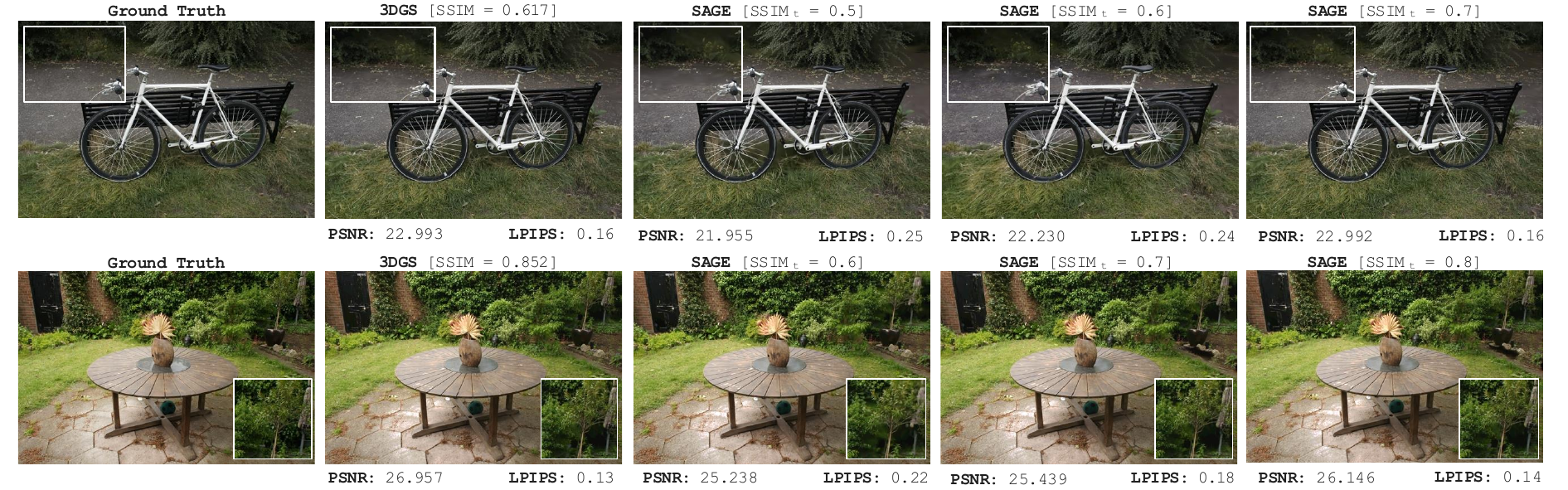} \vskip -1ex
    \caption{Qualitative cross-view results on ``bicycle" and cross-scene results on "garden".}
    \label{fig:qualitative-bicycle-garden} \vskip -1.5ex
\end{figure*}

Experiments were conducted on the Mip-NeRF360 dataset~\cite{mipnerf360} considering PSNR, SSIM, Learned Perceptual Image Patch Similarity (LPIPS)~\cite{lpips} as quality metrics, and the number of Gaussians (\# Gauss.) to characterize the quality and the complexity of the scene. SAGE was implemented on top of the original 3DGS framework~\cite{kerbl3Dgaussians}, with point clouds of Gaussians saved and analyzed at intervals of $1000$ iterations throughout the optimization process.
Table~\ref{avg-bicycle} presents the cross-view results for scene ``bicycle" averaged across all training images. The reported per-label evaluations allow us to verify that there exist significant class-dependent variations: \textit{sky-other-merged}\footnote{\textit{-merged} notation refers to labels which comprehend multiple related COCO~\cite{coco} classes mapped to a single class, following the standards of~\cite{deeplab2tensorflow}.} class consistently exhibits high SSIM values (above $0.8$) even at early iterations, while for others, SSIM values improve progressively. 
The overall behavior of SSIM versus iterations can be seen in Fig.~\ref{fig:ssim-iter} for scene ``bicycle".  
The category-related quality appears to be related to the variation of the content behavior within the same semantic category ({\eg}, \textit{sky-other-merged} is quite uniform with low-frequency content while \textit{grass-merged} presents different textures depending on the distance from the viewpoint).
Visual quality is also affected by the distance of each class from the camera. %
Using this model, we created an iteration predictor capable of estimating the level of detail necessary for each class to achieve the target $\text{SSIM}_t$.
Table~\ref{avg-garden} reports the SSIM values for different iterations for scene ``garden". Considerations similar to ``bicycle" can be made. The number of Gaussians and SSIM values evolve differently for different semantic categories over time, but interestingly, as for classes \textit{grass-merged} and \textit{pavement-merged} that appear both in ``bicycle" and ``garden",  SSIM metric increases proportionally with the number of Gaussians over time. A cross-scene evaluation for classes \textit{grass-merged}, \textit{pavement-merged} and \textit{tree-merged} (see highlighted rows in Tables~\ref{avg-bicycle} and~\ref{avg-garden}) showed that the model, trained on ``bicycle", could generalize to ``garden" with reasonable success. The SSIM values remained consistent across both scenes, indicating that SAGE can predict the appropriate iteration to achieve the target SSIM$_{t}$ in the new scene. However, some variations were observed in the prediction of the optimal iteration, suggesting that scene-dependent factors affect the precision of the model.

Fig.~\ref{fig:fitting} proves the model's predictive power, enabling adaptive reconstruction of scenes by integrating elements with varying LOD. Three targets SSIM are selected for SAGE evaluation ({\ie}, $0.5$, $0.6$, $0.7$ for scene ``bicycle") and tested on both the original training views and the novel synthesized perspectives, robustly assessing our method's capabilities. 
Our results show that the SSIM follows a distance-dependent trend, where closer objects undergo sharper SSIM variations. The proposed two-phase decay model (Eq.~\ref{eq:fitting}) captures this effect, prioritizing foreground detail while preventing excessive refinement of background content. We use intermediate steps of 3DGS optimization rather than the final optimized state to avoid fully optimizing 3DGS and account for memory and processing constraints in real-world XR applications. Halting the optimization at different points per object reduces redundancy while maintaining high perceptual quality.
Table~\ref{tab:view-dep} highlights SAGE's performance for a specific view ({\ie}, ``{\rm \texttt{DSC8719}}"), showcasing the trade-offs between visual fidelity and resource efficiency. Selected iterations for each semantic category at each target $\text{SSIM}_t$ are highlighted for $t=0.5, t=0.6, t=0.7$ in green, yellow, and orange, respectively (if the same iteration is selected for more than one $\text{SSIM}_t$, the values are highlighted with the color of the highest iteration).
Compared to the baseline 3DGS (where all categories are optimized for 30'000 iterations), SAGE reduces Gaussian count from 5.83M to 3.05M at SSIM$_t = 0.5$, lowering memory usage from 1.45GB to 756.2MB. 
By accounting for minimum distance in optimization, SAGE prevents spread classes like \textit{sky-other-merged} or \textit{tree-merged} from suffering reduced quality in close-up regions.
At a fixed occupancy/memory, SAGE achieves higher visual quality in rendering, as clearly highlighted by the LPIPS metric (Table~\ref{tab:occupancy}).
Finally, \cref{fig:qualitative-bicycle,fig:qualitative-bicycle-garden} compare SAGE's qualitative results to groundtruth views and standard 3DGS, for three targets $\text{SSIM}_t$. Quality spreads differently across diverse semantic categories; however, SAGE obtains similar quality for every category with $\text{SSIM}_t=0.7$ while consistently reducing the occupancy. PSNR and LPIPS results are also reported under each scene, showing they are aligned with the SSIM behavior.
Note that being SAGE designed to optimize resource allocation rather than maximize perceptual quality alone, it may not always outperform 3DGS in SSIM and LPIPS. However, it always maintains comparable visual quality with significantly lower memory usage, extremely important for real-time XR applications.

\section{Conclusions}
\label{sec:concl}
In this work, we presented SAGE, a novel strategy for optimizing the rendering of 3D scenes driven by semantics. Our approach effectively balances performance and resource usage, minimizing the number of Gaussians while maintaining a target visual quality, which is crucial for XR applications and other resource-constrained environments. Through experiments on the Mip-NeRF360 dataset, SAGE demonstrated not only to achieve significant memory savings compared to the 3DGS baseline but also to be able to effectively adapt the LOD of each semantic category within the 3D scene, based on their visual characteristics. By isolating each category and reconstructing the scene for a given SSIM value, SAGE ensures smooth rendering at reduced computational overhead.
The flexibility and resource efficiency of the proposed solution makes it a promising approach for large-scale 3D scene rendering, especially for real-time XR applications.

\bibliographystyle{IEEEbib}
\bibliography{eusipco2025.bib}

\begin{thebibliography}{10}

\bibitem{cap19:pcc}
F.~Capraro and S.~Milani,
\newblock ``Rendering-aware point cloud coding for mixed reality devices,''
\newblock in {\em Proc. IEEE ICIP}, 2019.

\bibitem{jav21:qual}
A.~Javaheri, C.~Brites, F.~Pereira, and J.~Ascenso,
\newblock ``Point cloud rendering after coding: Impacts on subjective and objective quality,''
\newblock {\em {IEEE} Trans. Multim.}, 2021.

\bibitem{Bil14:lod}
F.~Biljecki, H.~Ledoux, and J.~Stoter,
\newblock ``Redefining the level of detail for 3{D} models,''
\newblock {\em GIM International}, 2014.

\bibitem{Has21:3Dsimple}
J.~Hasselgren, J.~Munkberg, J.~Lehtinen, M.~Aittala, and S.~Laine,
\newblock ``Appearance-driven automatic 3{D} model simplification,''
\newblock in {\em Proc. EGSR}, 2021.

\bibitem{dos14:ar_cognitive}
J.~T. Doswell and A.~Skinner,
\newblock ``Augmenting human cognition with adaptive augmented reality,''
\newblock in {\em Foundations of Augmented Cognition. Advancing Human Performance and Decision-Making through Adaptive Systems}, 2014.

\bibitem{VAUGHAN201665}
N.~Vaughan, B.~Gabrys, and V.~N. Dubey,
\newblock ``An overview of self-adaptive technologies within virtual reality training,''
\newblock {\em Computer Science Review}, 2016.

\bibitem{4351598}
H.~Kim, Y.~Yoon, and H.~Park,
\newblock ``Adaptation method for level of detail ({LOD}) of 3{D}contents,''
\newblock in {\em Proc. IFIP NPC}, 2007.

\bibitem{camuffo2022deep}
E.~Camuffo, F.~Battisti, F.~Pham, and S.~Milani,
\newblock ``Deep 3d model optimization for immersive and interactive applications,''
\newblock in {\em Proc. IEEE EUVIP}, 2022.

\bibitem{nerf}
B.~Mildenhall, P.~P. Srinivasan, M.~Tancik, J.~T. Barron, R.~Ramamoorthi, and R.~Ng,
\newblock ``Nerf: representing scenes as neural radiance fields for view synthesis,''
\newblock {\em Commun. ACM}, 2021.

\bibitem{kerbl3Dgaussians}
B.~Kerbl, G.~Kopanas, T.~Leimk{\"u}hler, and G.~Drettakis,
\newblock ``3d gaussian splatting for real-time radiance field rendering,''
\newblock {\em ACM Transactions on Graphics}, 2023.

\bibitem{mipnerf360}
J.~T. Barron, B.~Mildenhall, D.~Verbin, P.~P. Srinivasan, and P.~Hedman,
\newblock ``Mip-nerf 360: Unbounded anti-aliased neural radiance fields,''
\newblock {\em Proc. IEEE CVPR}, 2022.

\bibitem{deeplabv2}
L.~Chen, G.~Papandreou, I.~Kokkinos, K.~Murphy, and A.~L Yuille,
\newblock ``Deeplab: Semantic image segmentation with deep convolutional nets, atrous convolution, and fully connected crfs,''
\newblock {\em IEEE Trans. Pattern Anal. Mach. Intell.}, 2017.

\bibitem{6178830}
O.~Oyman and S.~Singh,
\newblock ``Quality of experience for {HTTP} adaptive streaming services,''
\newblock {\em IEEE Communications Magazine}, 2012.

\bibitem{Wang17:qoe_has}
C.~Wang, D.~Bhat, A.~Rizk, and M.~Zink,
\newblock ``Design and analysis of {QoE}-aware quality adaptation for {DASH}: A spectrum-based approach,''
\newblock {\em ACM Trans. Multimedia Comput. Commun. Appl.}, 2017.

\bibitem{10.1145/3488660.3493802}
D.~Lorenzi, M.~Nguyen, F.~Tashtarian, S.~Milani, H.~Hellwagner, and C.~Timmerer,
\newblock ``Days of future past: an optimization-based adaptive bitrate algorithm over http/3,''
\newblock in {\em Proc. EPIC}, New York, NY, USA, 2021, EPIQ '21, p. 8–14, Association for Computing Machinery.

\bibitem{9122485}
T.~N. Duc, C.~T. Minh, T.~P. Xuan, and E.~Kamioka,
\newblock ``Convolutional neural networks for continuous {QoE} prediction in video streaming services,''
\newblock {\em IEEE Access}, 2020.

\bibitem{lightgaussian}
Z.~Fan, K.~Wang, K.~Wen, Z.~Zhu, D.~Xu, and Z.~Wang,
\newblock ``Lightgaussian: Unbounded 3d gaussian compression with 15x reduction and 200+ fps,'' 2023.

\bibitem{C3DGS}
J.~C. Lee, D.~Rho, X.~Sun, J.~H. Ko, and E.~Park,
\newblock ``Compact 3d gaussian representation for radiance field,''
\newblock in {\em Proc. IEEE CVPR}, 2024.

\bibitem{multiscale3dgaussiansplatting}
Z.~Yan, W.~F. Low, Y.~Chen, and G.~H. Lee,
\newblock ``Multi-scale 3d gaussian splatting for anti-aliased rendering,''
\newblock in {\em Proc. IEEE CVPR}, 2024.

\bibitem{octreegs}
K.~Ren, L.~Jiang, T.~Lu, M.~Yu, L.~Xu, Z.~Ni, and B.~Dai,
\newblock ``Octree-gs: Towards consistent real-time rendering with lod-structured 3d gaussians,'' 2024.

\bibitem{flod}
Yunji Seo, Young~Sun Choi, Hyun~Seung Son, and Youngjung Uh,
\newblock ``Flod: Integrating flexible level of detail into 3d gaussian splatting for customizable rendering,'' 2024.

\bibitem{campagnolo2023fully}
D.~Campagnolo, E.~Camuffo, U.~Michieli, P.~Borin, S.~Milani, and A.~Giordano,
\newblock ``Fully automated scan-to-bim via point cloud instance segmentation,''
\newblock in {\em Proc. IEEE ICIP}, 2023.

\bibitem{mari2023cactus}
D.~Mari, E.~Camuffo, and S.~Milani,
\newblock ``Cactus: Content-aware compression and transmission using semantics for automotive lidar data,''
\newblock {\em Sensors}, 2023.

\bibitem{schieber2024semanticscontrolledgaussiansplattingoutdoor}
H.~Schieber, J.~Young, T.~Langlotz, S.~Zollmann, and D.~Roth,
\newblock ``Semantics-controlled gaussian splatting for outdoor scene reconstruction and rendering in virtual reality,'' 2024.

\bibitem{lpips}
Richard Zhang, Phillip Isola, Alexei~A. Efros, Eli Shechtman, and Oliver Wang,
\newblock ``The unreasonable effectiveness of deep features as a perceptual metric,'' 2018.

\bibitem{coco}
Tsung-Yi Lin, Michael Maire, Serge Belongie, Lubomir Bourdev, Ross Girshick, James Hays, Pietro Perona, Deva Ramanan, C.~Lawrence Zitnick, and Piotr Dollár,
\newblock ``Microsoft coco: Common objects in context,'' 2015.

\bibitem{deeplab2tensorflow}
M.~Weber, H.~Wang, S.~Qiao, J.~Xie, M.~D. Collins, Y.~Zhu, L.~Yuan, D.~Kim, Q.~Yu, D.~Cremers, L.~Leal-Taixe, A.~L. Yuille, F.~Schroff, H.~Adam, and L.~Chen,
\newblock ``Deeplab2: A tensorflow library for deep labeling,'' 2021.

\end{thebibliography}

\end{document}